\documentclass[pre,aps,floatfix,preprint]{revtex4-1}

\usepackage{amsmath,amssymb}
\usepackage{graphicx}
\usepackage{psfrag}
\usepackage{graphicx}
\usepackage{color}
\usepackage[normalem]{ulem}
\usepackage{lipsum}

\def\beq{\begin{equation}}
\def\eeq{\end{equation}}
\def\bea{\begin{eqnarray}}
\def\eea{\end{eqnarray}}

\begin{document}
\title{When Hopf meets saddle: bifurcations in the diffusive Selkov model for glycolysis} 
 \author{Abhik Basu}\email{abhik.basu@saha.ac.in,abhik.123@gmail.com}
\affiliation{Condensed Matter Physics Division, Saha Institute of
Nuclear Physics, Calcutta 700064, West Bengal, India}
\author{Jayanta K Bhattacharjee}\email{jayanta.bhattacharjee@gmail.com}
\affiliation{School of Physical Sciences, Indian Association for the Cultivation of Sciences, 2A and 2B Raja S C Mullick 
Road, Calcutta 700032, West Bengal, India}

\date{\today}

\begin{abstract}
 We study the linear instabilities and bifurcations in the Selkov model for glycolysis with diffusion. We show that this model has a zero wave-vector, finite frequency Hopf bifurcation to a growing  oscillatory but spatially homogeneous state and a saddle-node bifurcation to a growing inhomogeneous state with a steady pattern with a finite wavevector. We further demonstrate that by tuning the relative diffusivity of the two concentrations, it is possible to make both the instabilities to occur at the same point in the parameter space, leading to an unusual type of {\em codimension-two} bifurcation. We then show that in the vicinity of this bifurcation the initial conditions decide whether a spatially uniform oscillatory or a spatially periodic steady pattern emerges in the long time limit. 
\end{abstract}

\maketitle

\section{Introduction}

The merging of a saddle node bifurcation and a Hopf bifurcation is a common feature of thermal~\cite{thermohaline-convec} and binary liquid convection~\cite{bin-liq-convec}. 
Convective instability (Rayleigh-Benard convection) occurs when a fluid is heated from below and occurs as a stationary instability (i.e., a transcritical bifurcation) (for classification of bifurcations see, e.g., Ref.~\cite{strogatz}) at a critical Rayleigh number in a parallel plate geometry. The plates are taken to be of linear dimension $L$ and separated by a distance $d$ in the vertical direction. The enclosed fluid is heated from below and a temperature difference $\Delta T$ is maintained between the plates. The Rayleigh number $\cal R$ is defined as 
\begin{equation}
{\cal R}=\frac{\alpha \Delta T g d^3}{\nu\lambda},\label{ray-no}
\end{equation}
where $g$ is the acceleration due to gravity, $\lambda$ is the thermal diffusivity and $\nu$ the kinematic viscosity. For $L/d\gg 1$ (large aspect ratio), convection sets in in the form of rolls of wavenumber $k$ ($\simeq 3.1/d$ at the threshold) at the critical Rayleigh number $R_c\approx 1708$. For $R>R_c$, one has steady convection (no time dependence). The bifurcation that occurs at $R=R_c$ is a saddle-node bifurcation where one eigenvalue of the stability matrix vanishes and subsequently becomes positive for $R>R_c$. The uncontrolled growth of the linear system is eventually arrested by nonlinearities in the hydrodynamic equations.

The situation changes dramatically if one uses a liquid mixture (i.e., a binary liquid) like water and alcohol for the study of the convective instability. The temperature gradient now brings in a concentration gradient, and in addition to the possibility of steady convection observed above, there is also a possibility of an oscillatory convection occurring via a Hopf bifurcation~\cite{binfluid-inst}. The threshold Rayleigh numbers, $R_s$ for steady convection and $R_o$ for oscillatory convection, are in general different. The observed instability is the one with the lower threshold.  However, by varying a parameter of the fluid (generally the Soret coefficient~\cite{soret}) which measures the response of the local concentrations of the two liquids to an imposed temperature gradient), one can actually set up a situation where $R_s=R_o$, i.e., where the saddle node and Hopf bifurcations meet~\cite{binfluid-inst}. The point is called a {\em codimension-two bifurcation}. Interestingly at the codimension-two point, the onset frequency of the Hopf bifurcation goes to zero. Further, in the binary mixture bifurcations, the wavenumbers for the periodic convection cells is taken to be the same for both stationary and oscillatory convections~\cite{binfluid-inst}. The convective state that is born at $R=R_s$ has the form $A(t) \cos {\bf k}_c\cdot {\bf x}$, where $k_c$ is a critical wavevector,  $\bf x$ is the in-plane coordinate, and $A(t)$ is any one of the physical variables, e.g., the velocity, temperature or concentration anove the convection threshold $R_s$, and is an exponentially growing function of time $t$ for $R>R_s$, i.e., $A(t)=C\exp(\epsilon_s t)$, where $C$ is a constant and $\epsilon_s \propto (R-R_s)/R_s$. On the other hand the  state produced at $R=R_o$, has the form $B(t) \exp(i\omega_0 t)\cos ({\bf k}_c\cdot {\bf x})$, where $B(t)=C'\exp(\epsilon_o t)$ with $C'$ a constant and $\epsilon_o\propto (R-R_o)/R_o$. The onset frequency $\omega_0$ follows from the linear stability analysis. Near the codimension two point where $R_s=R_o=R_c$ and $\omega_0=0$, we have the stationary convection solution going as $C\exp(\epsilon_s t)\cos ({\bf k}_c\cdot {\bf x})$ and the oscillatory convective solution going as $C'\exp[(\alpha + i\beta)\epsilon_o t]\cos ({\bf k}_c\cdot {\bf x}),\,\alpha>0$. The important point to note is that at the codimension two point where $\epsilon_s=0=\epsilon_o$, both solutions have the same structure. The different aspects of the convective instabilities in a binary fluid mixture have been examined in various forms by Silber and Knonloch~\cite{silber}, Knobloch and Moore~\cite{moore}, St. Hollinger and L\"ucke~\cite{holli} and F\"utterer~\cite{futt}; see also Refs.~\cite{gall,ricard} for recent general reviews on related problems.

In this article, we study a very different meeting of the Hopf and a saddle-node bifurcation in a reaction-diffusion system. In the original Selkov model for glycolysis~\cite{selkov}, one has two characteristic reaction rates $a$ and $b$, which define the parameter plane. The system allows a Hopf bifurcation with wavevector $k_c=0$ over a set of points $a_c,b_c$, where the concentration fields have the generic form $A({\bf X}) \exp (\lambda t)$, with ${\bf X}$ being the position coordinate, such that right at the bifurcation point $A({\bf X})$ is a constant. Further $\lambda =0$ at $a=a_c$ and $b=b_c$ while $\omega_0$ is a number of ${\cal O}(1)$. The system in the presence of diffusion also allows saddle-node bifurcation to a state with a steady pattern with a definite perodicity (hence, a finite wavevector $k_c$), over another set of points $a_c',b_c'$ in the immediate vicinity of which the concentration fields have the form $B(t)\cos ({\bf k}_c\cdot{\bf x})$, where $B(t)\sim \exp (\tilde \epsilon t)$ with $\tilde \epsilon = \alpha (a-a_c') + \beta (b-b_c')$. The codimension-two point in the present scenario occurs when $a_c=a_c'$ and $b_c=b_c'$. At this point the structure of the solution is of the generic form $A\cos \omega_0 t + B\cos {\bf k}_c \cdot {\bf x}$, which is very different from the binary liquid codimension two point where the solution is $\cos k_c x$ without any time-dependence. We address the structure of the bifurcation and the pattern formation in the vicinity of this unusual codimension two point in this work. A similar study on the codimension two point in the Brusselator model is available in Ref.~\cite{bruss}. The rest of this article is organized in the following manner. In Sec.~\ref{model}, we introduce the Selkov model for glycolysis with diffusion.
Then, in Sec.~\ref{lin-ins} we analyse the linear instabilities in the model and discuss the ensuing phase diagram in the parameter space. Next, in Sec.~\ref{lin-amp} we set up the amplitude equations. Finally, in Sec.~\ref{conclu} we summarise and conclude.

\section{Selkov model for glycolysis}\label{model}

The Selkov model for glycolysis was introduced to model glycolytic oscillations and has two species. The model equations read
\begin{eqnarray}
 \frac{\partial \rho_1}{\partial t} &=&-\rho_1 + a\rho_2 + \rho_1^2\rho_2 + \nabla^2 \rho_1,\label{mod1}\\
 \frac{\partial \rho_2}{\partial t} &=&  b - a\rho_2 -\rho_1^2 \rho_2 + D\nabla^2 \rho_2,\label{mod2}
\end{eqnarray}
where $\rho_1,\,\rho_2$ are the dimensionless concentrations of ADP (adenosine diphosphate)  and
F6P (fructose-6-phosphate), respectively~\cite{selkov}. We have added diffusion 
terms $\nabla^2\rho_1$ and $D\nabla^2 \rho_2$ in (\ref{mod1}) and (\ref{mod2}) respectively
that represent diffusion of the two species in space; the conventional Selkov model does
not consider diffusion~\cite{selkov}. All the parameters $a,b,D$ are positive. 
Notice that without diffusion, (\ref{mod1}) and (\ref{mod2}) are just two coupled ordinary 
differential equations (ODEs) that define a dynamical system, where as with 
diffusion they become partial differential equations (PDEs).

\section{Linear instabilities}\label{lin-ins}

At the fixed points of the model equations (\ref{mod1}) and (\ref{mod2}) $\rho_1$ and $\rho_2$ are constants given by
\begin{equation}
 \rho_1^*=b,\;\;\;\rho_2^*=\frac{b}{a+b^2}.\label{fps}
\end{equation}
Equations (\ref{mod1}) and (\ref{mod2}) may be linearised around the fixed points (\ref{fps}) to give
\begin{eqnarray}
 \frac{\partial u}{\partial t} &=& \frac{b^2-a}{b^2+a} u + av + b^2 v + \nabla^2 u,\label{modlin1} \\
 \frac{\partial v}{\partial t} &=& -(b^2+a)v - \frac{2ub}{b^2+a} + D\nabla^2 v,\label{modlin2}
\end{eqnarray}
where $u=\rho_1-\rho_1^*,\,v=\rho_2 - \rho_2^*$. Since (\ref{modlin1}) and (\ref{modlin2}) are PDEs, they actually correspond to an {\em infinite} number of modes, which may be conveniently labeled by the Fourier wavevector $\bf k$. The stability matrix $J$ for the pair of equations (\ref{modlin1}) and (\ref{modlin2}) take the form in the Fourier space
\begin{equation}
 J(k^2) = \left(\begin{array}{cc}
                 \frac{b^2 -a^2}{b^2+a^2} - k^2 & a+b^2\\
                 -\frac{2b^2}{a+b^2} & -a -b^2 - Dk^2
                \end{array}
\right).\label{j-matrix}
\end{equation}
The corresponding eigenvalues are given by
\begin{eqnarray}
 \lambda_\pm (k^2) &=&\frac{1}{2}[{\tt Tr}\pm \sqrt{{\tt Tr}^2 - 4{\tt Det}}],\label{eigen1}
\end{eqnarray}
where ${\tt Tr}$ and ${\tt Det}$, respectively, are the trace and determinant of the matrix $J(k^2)$, 
and both of these are functions of $k^2$. 
\begin{eqnarray}
 {\tt Tr}(k^2) &=& \frac{b^2-a}{b^2+a} - (a+b^2) - k^2 (D+1),\label{tr-eq1}\\
 {\tt Det}(k^2) &=& a+b^2 + k^2 (a+b) - Dk^2 \frac{b^2-a}{b^2+a} + Dk^4.\label{det-eq1}
\end{eqnarray}

Linear instability occurs when the real part of one or both the eigenvalues 
pass through zero. This can happen when either (i) {\tt Tr} $(k^2)$=0, 
when both $\lambda_\pm (k^2)$ become fully imaginary,  or (ii) {\tt Det} ($k^2$)=0, when 
$\lambda_-(k^2)$ entirely vanishes, for some $k$-values. 
The former is the condition for the onset of Hopf bifurcation, 
where as the second one is for saddle-node bifurcation. 

\begin{figure}[htb]
 \includegraphics[width=10cm]{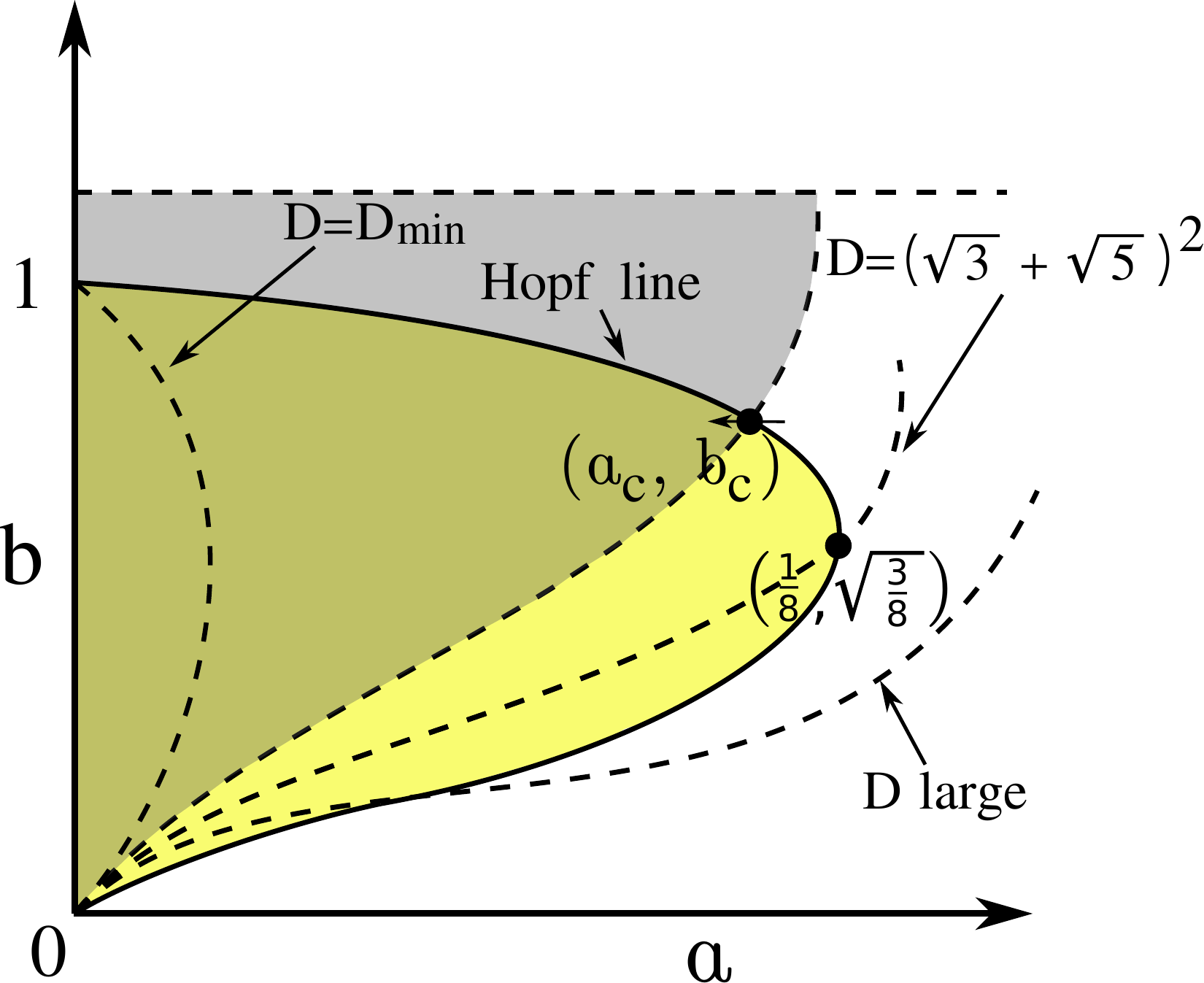}
 \caption{Schematic phase diagram of the diffusive Selkov model in the $(a-b)$ plane. The area enclosed between the continuous line and the axes (marked in yellow) is region that displays Hopf bifurcation to a uniform oscillatory state without diffusion. Different broken lines are boundaries of the saddle-node instabilities parametrised by $D$. For instance, the grey shaded region for some given $D$ corresponds to the saddle-node instabilities. The overlap of the yellow and grey shaded regions correspond to parameter values for which both Hopf bifurcation to an spatially uniform oscillatory state and saddle-node instabilities to a steady pattern are possible. The small arrow at the intersection between the Hopf boundary and the saddle-node instability boundary for some $D$ indicates formation of the instabilities as $a$ passes through $a_c$ from above for $b=b_c$ (see text).}\label{phase}
\end{figure}

It is clear from the form of $\lambda_\pm(k^2)$ that at the onset of Hopf bifurcation for 
the $k=0$ mode, all other modes with $k>0$ are {\em stable}. At $k=0$, at the onset $\lambda_\pm (k^2=0)$ 
are fully imaginary corresponding to a Hopf frequency $\omega_0 =\sqrt{a+b^2}$~\cite{strogatz}. 
This implies a steady oscillation permeating the entire system; the system remains spatially
homogeneous everywhere. The phase boundary in the $a-b$ plane that demarcates 
a steady homogeneous phase and a phase with oscillatory instability (i.e., with a growing
amplitude)is given by~\cite{strogatz} 
\begin{equation}
 {\tt Tr}(k^2=0)=0\implies b^2 = \frac{1}{2}(1-2a \pm\sqrt{1-8a}),\label{hopf-line}
\end{equation}
as shown in Fig.~\ref{phase}. At the onset of Hopf bifurcation, i.e., on the line 
(\ref{hopf-line}) in the $a-b$ plane, only the mode $\omega = \omega_0,\,k=0$ is marginal, all other modes decay. Different finite-$k$ modes also undergo Hopf bifurcation, at the onset of which the $k=0$ mode has the maximum growth rate. Thus, the $k=0$ mode is the most relevant mode for Hopf bifurcation in the linear stability analysis. Notice that this Hopf bifurcation exists for all $D$, simply because the $k=0$ mode, the dominant mode at the onset of Hopf bifurcation, is unaffected by the diffusivity.

Linear instability also arises when ${\tt Det} (k^2) =0$, at which point one of the  eigenvalues 
$\lambda_-(k^2)$  vanishes entirely for some $k$-value. This is the saddle-node bifurcation. 
In our model, the threshold for this instability is given by the condition 
${\tt Det}(k_c^2)=0$, where $k_c$ is a preferred wavevector, which can be obtained from the condition 
\begin{equation}
 \frac{\partial {\tt Det}}{\partial k^2}|_{k^2 = k_c^2} =0.
\end{equation}
We have 
\begin{equation}
\frac{\partial {\tt Det}}{\partial k^2}|_{k^2 = k_c^2} = 2D k_c^2 + \Gamma_1 =0 \implies k_c^2 = -\frac{\Gamma_1}{2D} >0\implies \Gamma_1 < 0,
\end{equation}
where $\Gamma_1 = a+b^2 - D(b^2-a)/(a+b^2) <0$ for $k_c^2 >0$. On the other hand, at the threshold of the saddle-node instability, 
\begin{equation}
 {\tt Det}(k_c^2) =0\implies [a+b^2 + D \frac{a-b^2}{a+b^2}]=4D (a+b^2).
\end{equation}
Together with the requirement of $k_c^2 >0$ in a steady pattern, we find
\begin{equation}
 a+b^2 + D\frac{a-b^2}{a+b^2} = -2\sqrt D\sqrt{a+b^2} \label{patterm-boun}
\end{equation}
as the phase boundary in the $a-b$ plane for a given $D$, separating a homogeneous phase and a steady pattern with $k_c$ as the preferred wavevector. This curve intersects the $b$-axis ($a=0$) at $b=\sqrt D (-1 +\sqrt 2)$. Furthermore, as $a\rightarrow 0$, $b\rightarrow 0$ on this curve, i.e., the curve passes arbitrarily close to the origin. In order to ascertain its behaviour near the origin, we assume 
\begin{equation}
 b^2=a+\Gamma a^\gamma,\;\;\gamma\neq 1, \label{ansatz1}
\end{equation}
as $a\rightarrow 0$. Substituting (\ref{ansatz1}) in (\ref{patterm-boun}), we find in the limit $a\rightarrow 0$
\begin{equation}
 -\frac{D\Gamma a^\gamma}{2a + \Gamma a^\gamma} = -2\sqrt D [2a + \Gamma a^\gamma]^{1/2}.
\end{equation}
This has no solution for $\gamma < 1$. For $\gamma >1$, we find
\begin{equation}
 D\Gamma a^{\gamma -1} = 4\sqrt 2 \sqrt D\sqrt a\implies \Gamma =\frac{4\sqrt 2}{D},\;\gamma = \frac{3}{2}.
\end{equation}
On the other hand the phase boundary (\ref{hopf-line}) between the stable homogeneous phase and oscillatory instability phase very close to the origin takes the form
\begin{equation}
 b^2 = a + {\cal O}(a^2),\;a\rightarrow 0.
\end{equation}
Thus the phase boundary (\ref{patterm-boun}) lies {\em above} the boundary (\ref{hopf-line}) very close to the origin.

The upper part of the Hopf line (\ref{hopf-line}) meets are $b$-axis ($a=0$) at $b=1$. Intersection of the pattern boundary (\ref{patterm-boun}) with the $b$-axis depends upon $D$. The threshold value of $D$ for which (\ref{patterm-boun}) intersects the $b$-axis as well as (\ref{hopf-line}) at $(0,1)$ is given by
\begin{equation}
 D_{min}=\frac{1}{(\sqrt 2 -1)^2}\approx 5.83.\label{dmin}
 \end{equation}
For $D< D_{min}$, (\ref{patterm-boun}) never intersects (\ref{hopf-line}); for $D<D_{min}$ (\ref{patterm-boun}) intersects (\ref{hopf-line}) at $a>0,\,b<1$. For instance, the two branches 
of the Hopf boundary meet at $a=1/8,\,b=\sqrt{3/8}$. The pattern boundary passes through this point for $D=(\sqrt 2 + \sqrt 3)^2\approx 9.9> D_{min}$. In general, the point of intersection $(a_c,b_c)$ between the two lines is given by
\begin{equation}
 a_c=\frac{2D}{(D-1)^2} - \frac{8D^2}{(D-1)^4},\;b_c^2 = \frac{2D}{(D -1)^2} + \frac{8D^2}{(D-1)^4}, \label{ac-bc}
\end{equation}
which are parametrised by $D$. Thus, 
by varying $D>D_{min}$ the point of the intersection of (\ref{patterm-boun}) with (\ref{hopf-line}) can be continuously shifted. In the limit of $D\rightarrow\infty$ (\ref{ac-bc}) gives
\begin{equation}
 a_c=\frac{2}{D},\;b_c^2 =\frac{2}{D}.\label{ac-bc-infty}
\end{equation}
Hence, for very large $D$, $(a_c,b_c)\rightarrow (0,0)$. Further, by using (\ref{ac-bc}) we obtain
\begin{equation}
 k_c^2 = \frac{1}{\sqrt D} \sqrt{a_c+b_c^2}=\frac{2}{D-1},\;\omega_0 = \sqrt{a_c+b_c^2} = \frac{2\sqrt D}{D-1} \label{crit-values}
\end{equation}
at the point of intersection $(a_c,b_c)$. Furthermore, $u,\,v\sim \exp(\pm i\omega_0 t)$ and $u,\,v \sim \exp (\pm i {\bf k}_c\cdot {\bf x})$ are the solutions of (\ref{modlin1}) and (\ref{modlin2}), and are the  dominant modes at $(a_c,b_c)$; all other modes decay in time. Thus the general solutions of $u,\,v$ at $(a_c,b_c)$ must be linear combinations of $\exp(\pm i\omega_0 t)$ and $\exp (\pm i {\bf k}_c\cdot {\bf x})$  (see below for explicit forms) 
which are neither travelling nor standing waves, rather an oscillation superposed on a steady
pattern. 

\begin{figure}[htb]
 \includegraphics[width=10cm]{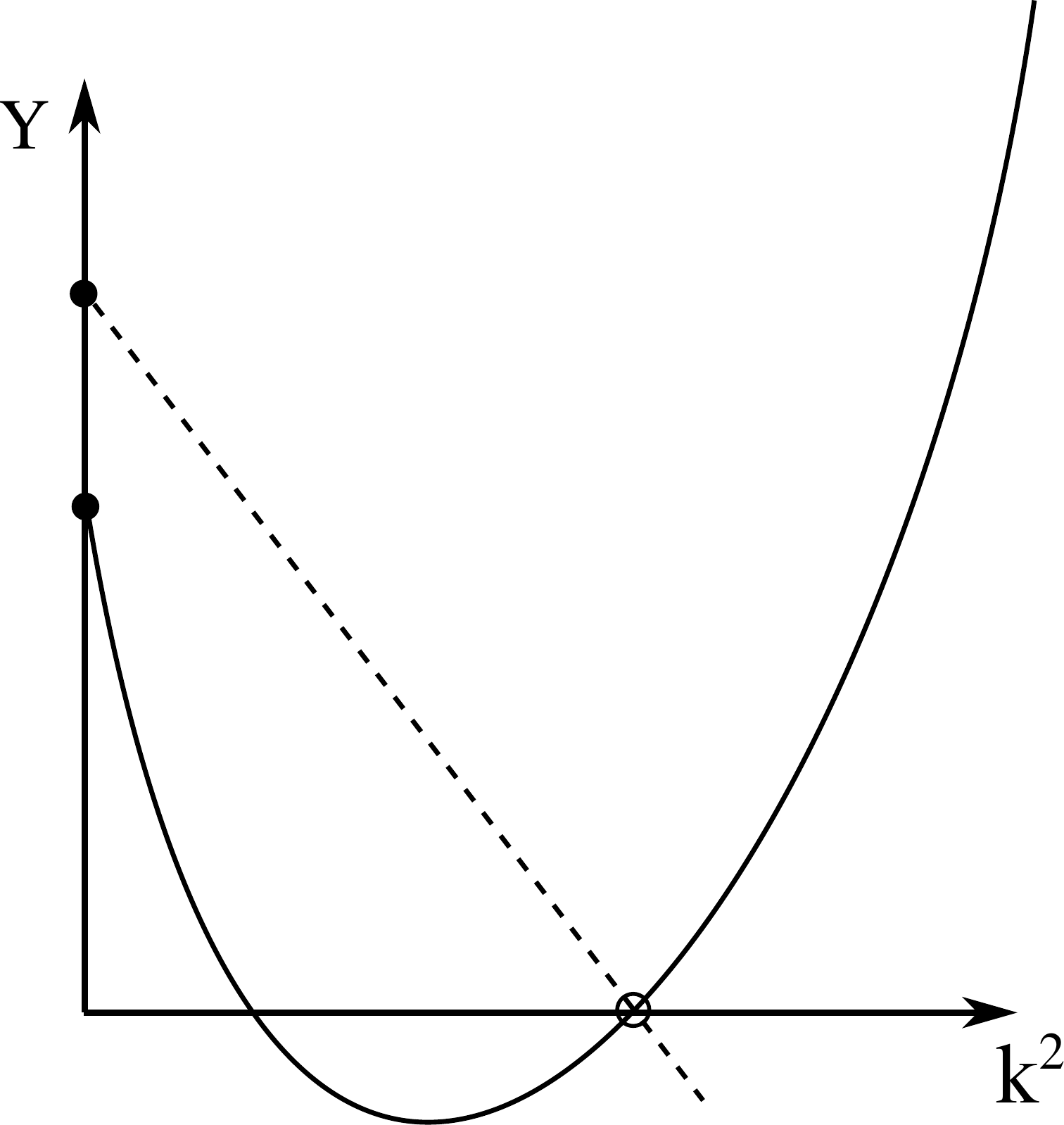}
 \caption{Possible origin of a Takens-Bogdanov bifurcation in the diffusive Selkov model. $Y$ in the y-axis refers to either ${\tt Det} (k^2)$ or ${\tt Tr}(k^2)$. The continuous line is the plot of (\ref{det-eq1}) and the broken line is the plot of (\ref{tr-eq1}); see text. Their meeting point is the small circle on the $k^2$-axis, which is the location of Takens-Bogdanov bifurcation in this model. }\label{takens}
\end{figure}

 Equations (\ref{tr-eq1}) and (\ref{det-eq1}) further suggest that in the diffusive Selkov model the threshold of a finite wavevector Hopf bifurcation can coincide with the threshold of a saddle-node (pattern) instability having a periodicity corresponding to the finite wavevector of the Hopf bifurcation; see Fig.~\ref{takens}. This is known as the Takens-Bogdanov bifurcation~\cite{takens}. We do not discuss it here further.

\section{Amplitude equations} \label{lin-amp}

At $(a_c,\,b_c)$ the amplitudes of the two modes are constants. Slightly away from $(a_c,\,b_c)$ and on the unstable side, these amplitudes grow exponentially in time. Let us set $b=b_c,\,a=a_c -\epsilon$, where $\epsilon$ is the {\em distance} from the threshold $(a_c,b_c)$, and is assumed to be small. At the threshold ($\epsilon=0$), only the modes with $\omega=\omega_0,\,k=0$ and $\omega=0,\,k=k_c$ survive and are marginal; all other modes decay. Thus at $\epsilon=0$, we can write
\begin{eqnarray}
u&=&A_1 \exp(i\omega_0t)+A_2\exp(i{\bf k}_c\cdot{\bf x}) + cc,\\
v&=&B_1\exp(i\omega_0t)+B_2\exp(i{\bf k}_c\cdot{\bf x}) + cc, \label{threshold1exp}
\end{eqnarray}
 where the direction of ${\bf k}_c$ is arbitrary; cc implies complex conjugates.

For $\epsilon>0$, the system gets unstable, and the modes should grow in time. 

In the linear theory, we find
\begin{eqnarray}
&&\left[\frac{\partial }{\partial t} + i\omega_0\right]A_1 =\frac{b^2-a_c}{a_c+b^2}A_1 + (a_c + b^2) B_1 + \epsilon A_1 \frac{2b^2}{(a_c+ b^2)^2} - \epsilon B_1 + \nabla^2 A_1.
\end{eqnarray}
At the threshold of the instability $(\epsilon=0)$, amplitudes $A_1$ and $B_1$ are related by
\begin{equation}
 A_1\left[i\omega_0 - \frac{b^2-a_c}{a_c+b^2}\right]= (a_c +b^2)B_1.
\end{equation}
Eliminating $B_1$, we obtain
\begin{eqnarray}
&& \frac{\partial A_1}{\partial t} = \epsilon A_1 \frac{2b^2}{(a_c+b^2)^2} + \frac{\epsilon A_1} {a_c+b^2} \left(i\omega_0 - \frac{b^2-a_c}{a_c+b^2}\right) + \nabla^2 A_1.
\end{eqnarray}
Similarly, for the pattern mode
\begin{eqnarray}
 &&\frac{\partial A_2}{\partial t}=\frac{b^2-a_c}{a_c+b^2}A_2 + \frac{\epsilon}{a_c+b^2}A_2 + (a_c+b^2)B_2-\epsilon B_2 -k_c^2 A_2 + 2i{\bf k}_c\cdot {\boldsymbol\nabla}A_2+ \nabla^2 A_2.
\end{eqnarray}
At the threshold of the instability $(\epsilon=0)$, amplitudes $A_2$ and $B_2$ are related by
\begin{equation}
 \frac{b^2-a_c}{a_c+b^2}A_2 - k_c^2 A_2 = -(a_c+b^2)B_2.
\end{equation}
Then eliminating $B_2$, we find
\begin{eqnarray}
 &&\frac{\partial A_2}{\partial t}=\epsilon A_2\frac{2b^2}{(a_c+b^2)^2} + \frac{\epsilon}{a_c+b^2} \left[\frac{b^2-a_c}{a_c+b^2} - k_c^2\right]A_2 + 2i{\bf k}_c\cdot {\boldsymbol\nabla}A_2+ \nabla^2 A_2.
\end{eqnarray}

Unsurprisingly, both $A_1$ and $A_2$ (and hence $B_1$ and $B_2$) grow exponentially in time. We now consider the nonlinear effects that eventually lead to saturation of the amplitudes in the long time limit~\cite{foot1}.  We start by expanding the model equations (\ref{mod1}) and (\ref{mod2}) about the fixed points (\ref{fps}) up to the cubic orders. Truncation at the cubic order is justified for small $\epsilon$. We find
 \begin{eqnarray}
  \frac{\partial u}{\partial t} &=& u \frac{b^2-a}{b^2 + a} + (a+b^2) v + \nabla^2 u  + N_u,\label{modnonlin1}\\
  \frac{\partial v}{\partial t} & = & - (a+b^2) v - 2u \frac{b^2}{a+b^2} + D\nabla^2 v + N_v,\label{modnonlin2}
 \end{eqnarray}
where $N_u$ and $N_v$ are the nonlinear terms:
\begin{equation}
 N_u = 2buv + \frac{u^2b}{a+b^2} + u^2 v = -N_v\equiv N,
\end{equation}
retaining up to the cubic contributions. In order to obtain the nonlinear amplitude equations, we again use the expansion (\ref{threshold1exp}), now with the understanding that the coefficients $A_1, A_2, B_1$ and $B_2$ are slowly varying functions of $\bf x$ and $t$. We substitute (\ref{threshold1exp}) for $u$ and $v$ in (\ref{modnonlin1}) and (\ref{modnonlin2}), and separately extract the coefficients of $\exp(i\omega_0 t)$ and $\exp(i{\bf k}_c\cdot {\bf x})$. The Hopf mode amplitude follows the nonlinear equation
\begin{eqnarray}
 \frac{\partial A_1}{\partial t}&=&\epsilon \frac{(D^2+1)^2 - 36D^2}{2(D-1)^4} A_1-\epsilon A_1\frac{i}{\omega_0} - (4+\frac{2}{D})A_1 |A_2|^2 - 3A_1|A_1|^2 \nonumber \\ &+& A_1|A_1|^2\frac{i}{\omega_0}+\nabla^2 A_1. \label{a1eq}
\end{eqnarray}
Similarly, the pattern mode amplitude follows the nonlinear equation
\begin{eqnarray}
 \frac{\partial A_2}{\partial t}&=& \epsilon a_c A_2 \frac{D^2+1-6D}{2(D-1)^4} (D^2+3) -(5+\frac{1}{D})A_2|A_1|^2-\frac{3}{2}(1+\frac{1}{D}) A_2|A_2|^2 \nonumber \\&+&2i{\bf k}_c\cdot {\boldsymbol\nabla}A_2+ \nabla^2 A_2. \label{a2eq}
\end{eqnarray}
Notice that while the coefficients of $A_1$ in (\ref{a1eq}) are in general complex, all the coefficients of $A_2$ in (\ref{a2eq}) are fully real. We thus set $A_2$ to be real with $|A_2|^2 = A_2^2$. In order to proceed further, we ignore the spatial dependences of $A_1$ and $A_2$; this assumption in effect reduces (\ref{a1eq}) and (\ref{a2eq}) just coupled ordinary differential equations. We further write $A_1=R\exp(i\phi)$ where $R$ is the magnitude and $\phi$ is the phase of $A_1$. It is easy to find the equation of motion for $R$, which reads
\begin{eqnarray}
 &&\frac{\partial R}{\partial t} = \epsilon \frac{(D^2+1)^2 - 36D^2}{2(D-1)^4}R - (4+\frac{2}{D})R A_2^2 -3R^3.\label{req}  
\end{eqnarray}
Similarly $A_2$ satisfies the ODE
\begin{eqnarray}
 \frac{\partial A_2}{\partial t}&=& \epsilon  A_2 \frac{D^2+1-6D}{2(D-1)^4} (D^2+3) -(5+\frac{1}{D})A_2R^2-\frac{3}{2}(1+\frac{1}{D}) A_2^3. \label{a2eq2}
\end{eqnarray}
As we show below, there is no jump in the order parameters $A_2$ and $R$ at the instability threshold $\epsilon =0$, as is clearly seen from (\ref{a2eq2}) and (\ref{req}). Thus, the bifurcations are {\em always forward}. At the fixed point, $\partial A_2/\partial t =0=\partial R/\partial t$. This gives four sets of fixed points $(R^*,A_2^*)$ that we obtain below as well as the linear stability of small fluctuations $\delta R$ and $\delta A_2$ around these fixed points (FP):

(i) FP1: $(R^*=0,A_2^*=0)$ together with
\begin{eqnarray}
 \partial_t \delta R&=& \epsilon \frac{(D^2+1)^2 - 36D^2}{2(D-1)^4}R,\\
 \partial_t \delta A_2 &=& \epsilon  A_2 \frac{D^2+1-6D}{2(D-1)^4} (D^2+3).
\end{eqnarray}
Hence, this fixed point is unstable in both $A_2$ and $R$ directions.


(ii)FP2: $ (R^*=0,\,{A_2^*}^2=\epsilon\frac{D^2+1-6D}{2(D-1)^4}(D^2+3)\frac{2D}{3(1+D)}$. Linear stability analysis gives
\begin{equation}
 \partial_t \delta R= \left[\epsilon \frac{(D^2+1)^2 - 36D^2}{2(D-1)^4} - (4+\frac{2}{D})A_2*^2\right],
\end{equation}
and
\begin{equation}
 \partial_t \delta A_2 = -2\epsilon  A_2 \frac{D^2+1-6D}{2(D-1)^4} (D^2+3).
\end{equation}
This fixed point is obviously stable along the $A_2$ direction. The stability along the $R$-direction is controlled by the sign of
\begin{equation}
 \Delta_R=(D+1)^2 - 36D^2 -\frac{2(2+4D)}{3(1+D)} (D^2+1-6D)(D^2+3).
\end{equation}
In the limit of $D\rightarrow \infty$ (i.e., when the species $v$ diffuses infinitely faster than species $u$), $\Delta_R=D^4 - 8D^4/3=-5D^4/3<0$ indicating stability along the $R$-direction as well. For finite $D$, we have evaluated $\Delta_R$ numerically and found it be negative for $D_{min}\leq D \leq 10^6$, suggesting that the fixed point $ (R^*=0,\,{A_2^*}^2=\epsilon\frac{D^2+1-6D}{2(D-1)^4}(D^2+3)\frac{2D}{3(1+D)}$ is globally linearly stable for all $D>D_{min}$.

(iii) FP3: $({R^*}^2=\epsilon \frac{(D^2+1)^2 - 36 D^2}{6(D-1)^4}, A_2=0)$. Linear stability analysis gives
\begin{equation}
 \partial_t \delta R= -2\left[\epsilon \frac{(D^2+1)^2 - 36D^2}{2(D-1)^4} \right]\delta R,
\end{equation}
implying stability along the $R$-direction. Further,
\begin{equation}
 \partial_t \delta A_2 = \left[\epsilon \frac{D^2+1-6D}{2(D-1)^4}(D^2+3) - \frac{5D+1}{D} {R^*}^2\right] \delta A_2.
\end{equation}
Thus, the stability along the $A_2$-direction is controlled by the sign of 
\begin{equation}
 \Delta_A=(D^2+1-6D)(D^2+3)-(5D+1)[(D^2+1)^2 - 36D^2]\frac{1}{3D}.
\end{equation}
In the limit of $D\rightarrow\infty$, $\Delta_A=D^4-5D^4/3=-2D^4/3<0$, implying stability. For finite values of $D$, we have evaluated $\Delta_A$ numerically and found it to be negative for $D_{min}\leq D \leq 10^6$, suggesting that the fixed point is linearly stable. Thus, the fixed point  $({R^*}^2=\epsilon \frac{(D^2+1)^2 - 36 D^2}{6(D-1)^4}, A_2=0)$ should be linearly stable for all values of $D\geq D_{min}$.


(iv) FP4: Both ${R^*}^2> 0,\,{A_2^*}^2 >0$. The solutions are written as
\begin{eqnarray}
 {R^*}^2&=&\epsilon\frac{\gamma_1\alpha_2-\gamma_2\alpha_1}{\beta_1\alpha_2-\beta_2\alpha_1},\\
 {A_2^*}^2&=&=\epsilon\frac{\beta_1\gamma_2-\beta_2\gamma_1}{\beta_1\alpha_2-\beta_2\alpha_1},
\end{eqnarray}
where 
\begin{eqnarray}
 &&\gamma_1=\frac{(D^2+1)^2-36D^2}{2(D-1)^4},\gamma_2 = \frac{D^2+1-6D}{2(D-1)^4}(D^2+3), \\&&\alpha_1= 4+\frac{2}{D},\alpha_2=5+\frac{1}{D},\beta_1=3,\beta_2=\frac{3}{2}\left(1+\frac{1}{D}\right).
\end{eqnarray}

For physically acceptable solutions, we must have ${A_2^*}^2>0,\,{R^*}^2>0$. In the limit of $D\rightarrow \infty$, ${R^*}^2=\epsilon/18$ and ${A_2^*}^2= \epsilon/12$, making these admissible solutions. At finite $D$, the solutions are numerically found to be positive for $D_{min}\le D\leq 10^6$.

We now look for the linear stability of these solutions. We find
\begin{eqnarray}
 \partial_t \delta R &=&  \left[ \epsilon\frac{(D^2+1)^2 - 36D^2}{2(D-1)^4} - (4+\frac{2}{D}){A_2^*}^2 -9{R^*}^2\right]\delta R -2  (4+\frac{2}{D}) {A_2^*}{R^*}\delta A_2,\\
 \partial_t \delta A_2 &=& \left[\epsilon \frac{D^2+1-6D}{2(D-1)^4}(D^2+3) - \frac{5D+1}{D}{R^*}^2 - \frac{9}{2}(1+\frac{1}{D}){A_2^*}^2\right]\delta A_2 - 2\frac{5D+1}{D}R^*A_2^*\delta R.
\end{eqnarray}
We find that for any $D>D_{min}$, one of the eigenvalues is positive, making this fixed point linearly unstable.

\begin{figure}[htb]
\includegraphics[width=8cm]{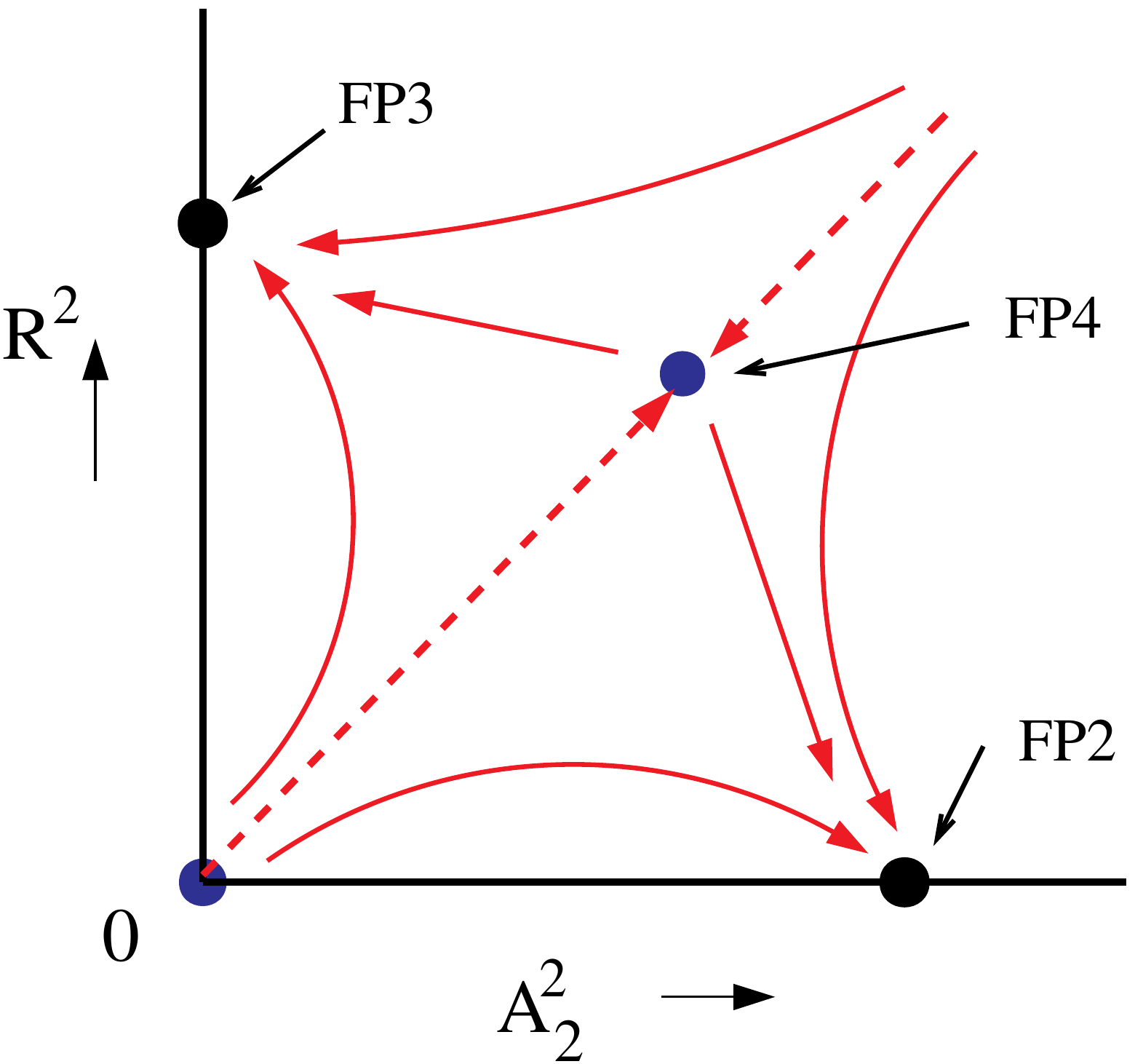}
 \caption{Flow lines around the fixed points in the $A_2^2-R^2$ plane. Filled blue circles represent linearly unstable fixed points and filled black circles represent linearly stable fixed points. The broken red line is the separatrix. Arrows denote directions of the flows.}\label{flow}
\end{figure}

The flow diagram around the fixed points are shown in Fig.~\ref{flow}. The equation of the separatrix that separates the basin of attractions of FP2 and FP3 is given by the condition
\begin{equation}
 {R^*}^2({\beta_1\gamma_2-\beta_2\gamma_1})= {A_2^*}^2 ({\gamma_1\alpha_2-\gamma_2\alpha_1}),
\end{equation}
which is unsurprisingly a straight line in the ${A_2^*}^2-{R^*}^2$ plane, passing through the origin; the slope $m$ of the separatrix is 
\begin{equation}
m=(\gamma_1\alpha_2-\gamma_2\alpha_1)/(\beta_1\gamma_2-\beta_2\gamma_1)= 2\frac{(5D+1) (D^2+1+6D) - (4D+2)(D^2+3)}{6(D^2+3)D - 3(D+1) (D^2+1+6D)}
\end{equation}
that clearly depends upon $D$. As $D\rightarrow\infty$ slope $m\rightarrow 2/3$.

Initial conditions lying below the separatrix flow towards FP3, where as those lying above flow towards FP2.
Thus, the precise initial conditions determine the ensuing final states for small $\epsilon$ near $(a_c,b_c)$, which is either a uniform state with oscillation (Hopf state controlled by FP2), or a steady pattern (controlled by FP3). When the system is controlled by FP2, the eventual final state should display spirals, targets etc~\cite{cross,arijit,dsr}, where as when it is controlled by FP3, the system should display steady patterns of a given periodicity~\cite{schneider,thesis}. 
This opens the question what one might observe if one crosses the boundaries away from $(a_c,b_c)$. In this case, one either crosses the boundary of the Hopf bifurcation first, or the saddle-node instability first. Consider the case, when one crosses the boundary of the saddle-node instability first. Upon crossing this boundary and {\em before} crossing the Hopf bifurcation boundary, the state is a patterned state with a given periodicity or a wavevector. At the boundary of the Hopf bifurcation, this state actually does not undergo an instability, for only a uniform state undergoes a Hopf bifurcation at this boundary. Similarly, if one crosses the Hopf bifurcation boundary first, a uniform oscillatory instability sets in. Upon meeting the saddle-node instability boundary, this oscillatory state does not undergo another instability as at the saddle-node instability boundary only a non-oscillatory uniform state undergoes an instability. Thus, depending upon which boundary the system meets first starting from a uniform state, a particular final state will be generated. For $D<D_{min}$ as one approaches from the uniform steady state, one necessarily meets the Hopf bifurcation boundary leading to a Hopf bifurcation to a uniform oscillatory state; for $D<D_{min}$ there are no patterned states. Of course, very far from the boundaries and near to the origin, there can be further instabilities of period or time scale doubling type, leading ultimately to spatio-temporal chaos. We do not discuss this here.

 \section{Summary and outlook}\label{conclu}
 
 We have developed a generic description for forward bifurcations near a co-dimension two point. To this end, we have studied the Selkov model for glycolysis with diffusion. Linear stability analysis is used to show that the model equations admit two independent linear instabilities - (i) a zero wavevector Hopf bifurcation from a uniform state to a uniform oscillatory state, and (ii) a finite wavevector saddle-node instability from a uniform steady state to a patterned state at zero frequency. We obtain the phase diagram spanned by the two model parameters. The thresholds of these two instabilities can be made to superpose on the same point in the phase diagram by tuning the diffusion constant $D$, which is a co-dimension two point. We have asked what the nature of the final state is very close to the threshold. To analyse this, we have set up the lowest order nonlinear amplitude equations for the Hopf and pattern modes, which are coupled by the nonlinear effects. We then show that the amplitude equations admit four the fixed points, all describing only forward bifurcations. Only  two of these are globally stable, with one corresponding to a uniform state with oscillation and the other to a steady pattern. Thus, depending upon the initial conditions, very close to the common instability threshold the model is to undergo either a Hopf bifurcation akin to the model without diffusion, or a saddle-node bifurcation, with no trace of the other being observed in experiments on representative physical systems. These results could also be verified by numerically solving the model partial differential equations. Our results are expected to be generic and should hold for any pair of amplitude equations having similar structure. While setting up the amplitude equations, we have neglected the higher order coefficients. This may be justified on the ground that the lowest order nonlinear terms give for all the amplitudes $A_1, A_2, B_1, B_2$ to be ${\cal O}(\epsilon)$. Any higher order nonlinear contributions (which are neglected here) would produce higher order in $\epsilon$ corrections to the amplitudes. Near the threshold, $\epsilon$ is small and hence those contributions from the higher order nonlinearities can be ignored here.

 \section{Acknowledgement}
One of us (A.B.) thanks the Alexander von Humboldt Stiftung, 
Germany for 
partial 
financial support 
through the Research Group Linkage Programme (2016).


\begin{thebibliography}{99}
\bibitem{foot1} At the same order in $\epsilon$, one would also have higher order modes with either frequency $2\omega_0$ (``frequency doubling''), or spatial modulations given by $2k_c$ (``wavevector doubling''), which get generated at the nonlinear orders. We ignored these modes here for simplicity.
\end{thebibliography}
\end{document}